\documentclass[aps,prl,twocolumn,superscriptaddress,floatfix,showpacm]{revtex4}
\usepackage{graphicx}
\usepackage{amssymb}
\usepackage{amsmath}
\usepackage{verbatim}

\ifx\pdftexversion\undefined
\usepackage[dvips]{hyperref}
\else
\usepackage{hyperref}
\fi
\hypersetup{
  colorlinks = true, linkcolor = blue
}

\begin{document}

\title{Competing structures in 2D-trapped dipolar gases}

\author{Barbara Gr\"anz}
\affiliation{Theoretische Physik, ETH Zurich, CH-8093
Zurich, Switzerland}
\author{Sergey E.\ Korshunov}
\affiliation{L.D.\ Landau Institute for Theoretical Physics, 
142432 Chernogolovka, Russia}
\author{Vadim B.\ Geshkenbein}
\affiliation{Theoretische Physik, ETH Zurich, CH-8093
Zurich, Switzerland}
\author{Gianni Blatter}
\affiliation{Theoretische Physik, ETH Zurich, CH-8093
Zurich, Switzerland}

\date{\today}

\begin{abstract}
We study a system of dipolar molecules confined in a two-dimensional trap and
subject to an optical square lattice. The repulsive long-range dipolar
interaction $D/r^3$ favors an equilateral triangular arrangement of the
molecules, which competes against the square symmetry of the underlying
optical lattice with lattice constant $b$ and amplitude $V$. We find the
minimal-energy states at the commensurate density $n = 1/b^2$ and establish
the complete square-to-triangular transformation pathway of the lattice with
decreasing $V$ involving period-doubled, solitonic, and distorted-triangular
configurations.
\end{abstract}


\maketitle

Competing structures and effects of commensuration appear in numerous physical
systems. Prominent examples are atoms on surfaces, e.g., Krypton on graphite
\cite{Specht_84}, vortices in modulated superconducting films
\cite{Daldini_74}, in periodic pinning arrays \cite{Tonomura_96}, and in a BEC
subject to an optical lattice \cite{Reijnders_04}, flux quanta in Josephson
junction arrays \cite{Webb_83}, or colloidal monolayers on periodic substrates
\cite{Leiderer_03}. A new realization of this physics is accomplished by
assembling cold dipolar molecules \cite{pol_mol} (e.g., KRb \cite{Wang_04} or
RbCs \cite{Sage_05}) in a two-dimensional (2D) optical trap and stabilizing
them with the help of a perpendicular electric field \cite{Buechler_07}.
Adding a square optical lattice provides an effective substrate potential
which competes against the triangular lattice arrangement favored by the
long-range repulsive dipolar interaction. As a result, the system is expected
to exhibit a variety of different configurations as a function of particle
density and strength of the substrate potential.  In this paper, we find the
minimal-energy states at commensurate density in the absence of quantum and
thermal fluctuations and thereby establish the complete transformation pathway
from the square to the triangular lattice.  Contrary to previous studies, the
cold molecule system, besides being clean, can be continuously tuned through
various configurations by changing system parameters such as particle density
and substrate potential amplitude. Even more, modifying the orientation or
number of lasers, the symmetry of the optical lattice can be changed.

\begin{figure}[h]
\begin{center}
\includegraphics[width=8cm]{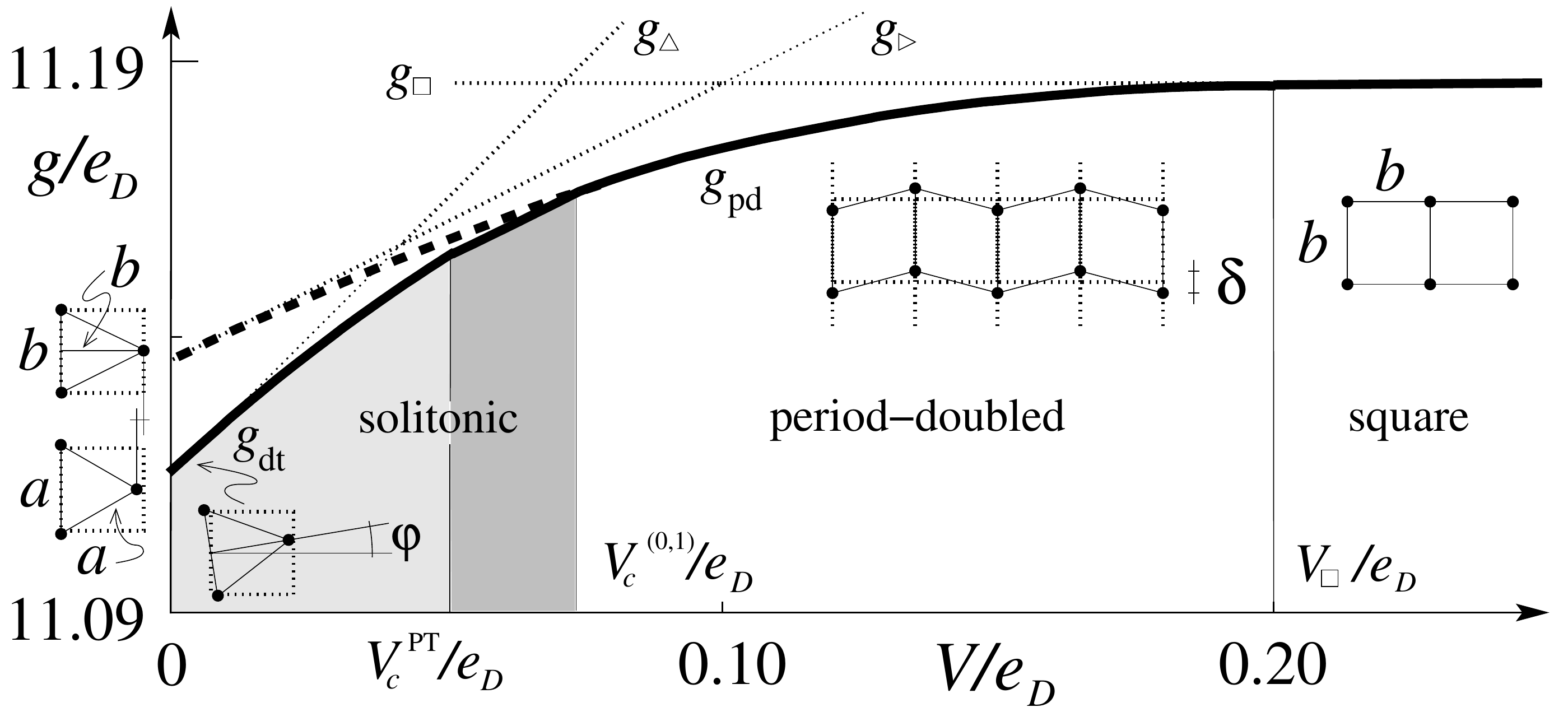}
\end{center}
\caption{\label{fig:phase_dia} Gibbs free energy of optimal states (thick
line), triangular at $V=0$, distorted and rotated triangular ($g_\mathrm{dt}$)
at small $V$, solitonic and period-doubled ($g_\mathrm{pd}$) at intermediate
$V$, and square for $V > V_{\scriptscriptstyle \square}$. Below the critical
potential $V_c^{\scriptscriptstyle (0,1)}$, the period-doubled phase smoothly
transforms into the triangular lattice via two soliton transitions involving
different soliton arrays. The dashed line extrapolates the energy
$g_\mathrm{pd}$ of the period-doubled phase. Dotted lines are energies of
rigid triangular ($\triangle$), isosceles ($\rhd$), and square ($\square$)
configurations.}
\end{figure}

In the simplest case, the transformation pathway between lattices with
different symmetries may involve a sequence of other uniform lattices.  An
interesting situation arises when new topological objects show up in
intermediate {\it non-uniform} phases.  The original `misfit problem' between
a particle lattice with lattice constant $a$ and a periodic substrate with
incommensurate periodicity $b\neq a$ has first been formulated in one
dimension (1D); these studies \cite{FrenkelKontorova_39,FrankVdMerwe_49} have
shown that the locked system at large potential $V$ (with particle separation
$b$) smoothly transforms into the free lattice (with separation $a$ between
particles) at $V=0$ via a non-uniform soliton phase, with soliton cores
approximating the free phase separating regions of locked phase. The
commensurate--incommensurate transition in the 2D analogue has been addressed
by Pokrovsky and Talapov \cite{PokrovskyTalapov_79,PT_pap,PT_book}; within
their `resonance approximation', the problem reduces to a 1D one and the
system develops a secondary structure in the form of a soliton-line array.
Going beyond the resonance approximation, we find that the
square-to-triangular transformation in the dipolar system involves three
separate transitions related to the formation of a period-doubled zig-zag
lattice as well as two instabilities towards non-uniform soliton phases, see
Fig.~\ref{fig:phase_dia}.

We consider a 2D-confined molecular gas with dipolar interaction $D/r^3$
between the molecules,
\begin{equation}\label{eq:Eint}
   E^\mathrm{int} = \frac{1}{2}\sum_{i\neq j} \frac{D}{r_{ij}^3},
\end{equation}
subject to an optical (substrate) lattice 
\begin{equation}\label{eq:Esub}
   E^\mathrm{sub} = \frac{V}{2}\sum_{i,\alpha} 
   \bigl[1-\cos({\bf q}_\alpha\!\cdot\! 
   {\bf r}_i)\bigr].
\end{equation}
The particles with density $n=1/b^2$ equal to commensurate filling (one
particle per minimum) are located at positions ${\bf r}_i$ with distances
$r_{ij} \equiv |{\bf r}_i - {\bf r}_j|$; the substrate potential involves
two modes ${\bf q}_1 = (q,0)$ and ${\bf q}_2 = (0,q)$ with $q=2\pi/b$.
Starting with the system's energy for $N$ particles trapped within the area
$A$, $E(A,N)= E^\mathrm{int} + E^\mathrm{sub}$, our task is to minimize the
Gibbs free energy per particle 
\begin{equation}\label{eq:gpN}
   g(p) = G(A,N)/N = [E(A,N)+pA]/N,
\end{equation}
where the thermodynamic limit with fixed density $n=N/A$ is implied. We choose
to work at fixed pressure $p$ rather than fixed chemical potential, as this
seems a better approximation to the experimental setup where molecules are
confined to a trap.  At $V = 0$, the molecules arrange in an equilateral
triangular lattice with a lattice constant $a = (4/3)^{1/4}\, b > b$ and
height $h = (3/4)^{1/2}\, a < b$; the resulting misfit parameter then is $s =
b/h-1 \approx 0.0746$. Given the purely repulsive interaction between
molecules, the density $n$ is related to the pressure $p = (3/2) n
e_{\scriptscriptstyle \triangle}$, with $e_{\scriptscriptstyle \triangle} =
e_{\scriptscriptstyle \triangle}^\mathrm{int} \approx
4.446\,e_{\scriptscriptstyle D}$ the interaction energy per particle in the
triangular lattice and $e_{\scriptscriptstyle D} = D/b^3$ the dipolar energy
(the prefactor is conveniently calculated with an Ewald summation technique
\cite{ewald_21}).  Upon switching on a small but finite potential $V>0$, the
rigid lattice assumes an energy $g_{\scriptscriptstyle \triangle}(V) =
e_{\scriptscriptstyle \triangle}(V)+p/n \approx 11.115 \,
e_{\scriptscriptstyle D} + V$, increasing with amplitude $V$ as each substrate
mode contributes with an average $V/2$ to the energy.  For a large potential
$V$, the molecules arrange in a square lattice with lattice constant $b < a$
and an energy $g_{\scriptscriptstyle \square} \approx 11.186\,
e_{\scriptscriptstyle D}$ independent of $V$ as all particles occupy potential
minima.  Besides the triangular and fully locked square lattices, a third
low-energy configuration \cite{Pogosov_03} is that of an isosceles triangular
lattice (below called the $bb$-lattice) with base $b$ and a height $b$ locked
to one substrate mode (we have to break the symmetry and choose the mode along
$x$) with an energy $g_{\scriptscriptstyle \rhd}(V)=11.136\,
e_{\scriptscriptstyle D}+V/2$.  The above expressions for
$g_{\scriptscriptstyle \triangle}(V)$, $g_{\scriptscriptstyle \rhd}(V)$, and
$g_{\scriptscriptstyle \square}$ already provide a reasonable approximation to
the energy $g$ versus potential $V$ diagram as illustrated in Fig.\
\ref{fig:phase_dia} (dotted lines).

Next, we account for deviations ${\bf u}_i$ of the particle coordinates ${\bf
r}_i = {\bf R}_i^\mathrm{latt}+{\bf u}_i$ from regular lattice positions ${\bf
R}_i^\mathrm{latt}$. Expanding Eq.\ (\ref{eq:Eint}) in the displacement field
${\bf u}_i$, the energy $g = g_\mathrm{latt} +\delta g$
picks up a term
\begin{equation}\label{eq:gint}
   \delta g^\mathrm{int} \approx \frac{1}{2N} \sum_{i,j} {{\bf u}_i}^T \,
   \hat{\Phi}^{\scriptscriptstyle D}
   ({\bf R}_{ij}^\mathrm{latt}) \,{\bf u}_j,
\end{equation}
with the elastic matrix $\hat{\Phi}^{\scriptscriptstyle D}({\bf
R}_{ij}^\mathrm{latt})$ depending on the chosen lattice; the substrate
potential contributes a second term $\delta e^\mathrm{sub}$ to $\delta g$,
$\delta g = \delta g^\mathrm{int} + \delta e^\mathrm{sub}$.

For a large substrate potential $V$, the substrate enforces a square lattice
with particle positions ${\bf R}_i^\mathrm{latt}={\bf R}_i^{\scriptscriptstyle
\square}$. Since the true configuration at $V=0$ is the {triangular} one, the
square lattice becomes unstable when {decreasing} $V$. The symmetry-breaking
instability is towards a period-doubled zig-zag phase \cite{Zhuralev_03} and
is conveniently analyzed in Fourier space; the elastic matrix
$\hat{\Phi}^{\scriptscriptstyle D} ({\bf k})$ exhibits negative eigenvalues,
with the most negative one $\phi^\perp_X = -3.958\, e_{\scriptscriptstyle D}
n$ located at the $X$ point $(\pi/b,0)$ of the Brillouin zone and describing a
shear distortion (alternatively, the symmetry breaking involves the point
$(0,\pi/b)$).  The contribution $\delta e^\mathrm{sub}$ shifts all eigenvalues
by $V q^2 /2$, thus stabilizing the square lattice.  The instability occurs
when the first eigenvalue crosses zero at
\begin{equation}\label{eq:V_sq}
   V_{\scriptscriptstyle\square} = -(2/q^2) \phi^\perp_X\approx
   0.201\, e_{\scriptscriptstyle D}.
\end{equation}
Decreasing $V$ below $V_{\scriptscriptstyle\square}$, the system transforms to
a period-doubled phase with two molecules per rectangular unit cell, see Fig.\
\ref{fig:phase_dia}, with lattice vectors ${\bf R}_1^\mathrm{rec}= (2b,0)$ and
${\bf R}_2^\mathrm{rec}= (0,b)$ and molecular positions ${\bf c}_1 = (0,u_1)$
and ${\bf c}_2 = (b,u_2)$ therein.  Inserting these coordinates into the
functional Eq.\ (\ref{eq:gpN}), we use the Poisson summation formula replacing
the real-space sum along $y$ by the reciprocal-space sum over $\ell q$ to
obtain the energy
\begin{eqnarray}\label{eq:eds}
   g_\mathrm{pd}(\sigma, \delta) \!&=&\! 8\pi e_{\scriptscriptstyle D}\!\!
   \sum_{i,\ell >0} \frac{\ell K_1[2\pi \ell (2i-1)]}{2i-1}\cos(q\ell\delta) \\
   \nonumber
   {\vspace{-4pt}}
   &&+(V/2) [1-\cos(q\sigma)\cos(q\delta/2)] + \mathrm{const.}
\end{eqnarray}
with $\sigma = (u_1+u_2)/2$ and $\delta = (u_1-u_2)$. The modified Bessel
function $K_1(z) \propto e^{-z}$ decays rapidly and we can limit the sum in
Eq.\ (\ref{eq:eds}) to the term $i=\ell=1$. Minimizing $g_\mathrm{pd}$ with
respect to $\delta$, we find that $\cos(q\delta/2) = (V/8\Delta)\cos(q\sigma)$
with $2\Delta = g_{\scriptscriptstyle\square} -g_{\scriptscriptstyle
\rhd}(V=0) \approx 0.0496 \, e_{\scriptscriptstyle D}$, and the energy reads
\begin{equation}\label{eq:g_pd}
   g_\mathrm{pd}(\sigma) = g_{\scriptscriptstyle \rhd}(V) -
   \frac{V^2}{32 \Delta} \cos^2(q \sigma).
\end{equation}
For the homogeneous period-doubled phase, $\sigma = 0$, i.e., the molecules
displace symmetrically around the substrate minima along $y$, and $\delta =
(b/\pi)\arccos(V/8\Delta)$. The condition $\delta = 0$ provides us with the
critical potential $V_{\scriptscriptstyle\square} = 8\Delta \approx 0.198\,
e_{\scriptscriptstyle D}$; this is close to the previous result
(\ref{eq:V_sq}), confirming that terms with $i>1$ or $\ell>1$ in Eq.\
(\ref{eq:eds}) are indeed small. The order parameter approaches zero as
$\delta \approx \pm (\sqrt{2}\, b/\pi) \sqrt{1-V/V_{\scriptscriptstyle
\square}}$, while $\delta = \pm b/2$ at $V=0$ describes the $bb$-lattice with
energy $g_{\scriptscriptstyle \rhd}$. The $\pm$ signs refer to the two
possibilities to break the symmetry when doubling the period, leading to twin
configurations with zig-zag structures shifted by $b$ along $x$. The
period-doubled phase then exists in four versions, with the zig-zag structure
manifest along $x$ or $y$, each with a twin shifted by $b$.  The energy
$g_\mathrm{pd}(V)$ of this phase resides below the energy
$g_{\scriptscriptstyle \rhd}(V)$ of the singly-locked isosceles phase, see
Fig.\ \ref{fig:phase_dia}.

Next, we focus our interest to weak substrate potentials $V$. The particle
coordinates then deviate from regular triangular lattice positions, i.e.,
${\bf R}_i^\mathrm{latt} ={\bf R}_i^{\scriptscriptstyle \triangle}$ and ${\bf
r}_i = {\bf R}_i^{\scriptscriptstyle \triangle} +{\bf u}_i$ in Eq.\
(\ref{eq:gint}).  For very small $V$, one can expand the substrate potential
to linear order in the displacement \cite{McTague_79} and minimize the
correction $\delta g$ in Fourier space. The force field involves the two modes
${\bf q}_\alpha$ of the substrate potential, folded back to the first
Brillouin cell of the particle lattice, ${\bf q}_\alpha - n_\alpha {\bf K}_1 -
m_\alpha {\bf K}_2 \equiv -{\bf p}_\alpha$,  with ${\bf K}_1$, ${\bf K}_2$ the
reciprocal lattice vectors of the (triangular) particle lattice,
$n_\alpha,~m_\alpha$ are appropriate integers, and we have included a minus
sign in the definition of ${\bf p}_\alpha$ for convenience.  The
minimal-energy configuration is found by rotating the triangular particle
lattice with respect to the square substrate potential and relaxing the
configuration in the force field. For a small misfit parameter $s$, one of the
vectors ${\bf p}_\alpha$ passes near zero, generating a large deformation (and
accordingly large energy gain) as the inverse elastic matrix
$[\hat{\Phi}^{\scriptscriptstyle D}]^{-1} ({\bf k} \to 0) \propto 1/k^2$ is
large at small ${\bf k}$.  Within the resonance approximation
\cite{PT_pap,PT_book}, only the dominant term in the relaxation deriving from
the small misfit vector, say ${\bf p}_1 = {\bf K}_1- {\bf q}_1$, is included,
while the small correction due to the other mode is dropped.  Within this
approximation, the optimal value of the angle $\varphi$ between the symmetry
axes of the particle lattice and the substrate (see Fig.\ \ref{fig:phase_dia})
is given by the same formula as derived by McTague and Novaco
\cite{McTague_79} for the accommodation of a triangular lattice on a substrate
with the same (triangular) symmetry but with a different lattice constant,
$\varphi_s \approx \sqrt{\nu} s$. Here, $\nu = (\kappa -\mu) /(\kappa+\mu)$ is
the Poisson ratio, with $\mu$ and $\kappa$ the shear and compression moduli.
For the dipolar interaction $\propto R^{-3}$, one has $\nu = 9/11$
\cite{PT_book} ($\kappa = 10 \mu$ and $\mu/n = (3/8)\, e_{\scriptscriptstyle
\triangle}$) and accordingly $\varphi = 3.86^\circ$.  The energy (to leading
order in $s$) of the distorted triangular phase reads
\begin{equation}\label{eq:dgpt}
   g_\mathrm{dt} = g_{\scriptscriptstyle \triangle} (V)
   - \frac{V^2}{64 s^2}\frac{n}{\mu} (1+\mu/\kappa)
\end{equation}
and we find a sinusoidal distortion field evolving along the direction ${\bf
z} \parallel {\bf p}_1$ enclosing an angle $\theta = \arctan\sqrt{\nu} \approx
42.13^\circ$ with the substrate lattice, i.e., near the diagonal.

With increasing $V$, this periodic distortion becomes large, of order $b$, and
turns into a soliton array as first described by Pokrovsky and Talapov
\cite{PT_pap,PT_book} within the same resonance approximation. Adopting a
continuum elastic description and retaining the full anharmonic form of the
substrate potential, they showed that the solution ${\bf u} =  {\bf
u}_\mathrm{g}+{\bf u}_\mathrm{p}$ minimizing the Gibbs free energy combines a
global deformation ${\bf u}_\mathrm{g}$ with a periodic modulation ${\bf
u}_\mathrm{p}$ that accounts for the soliton array. The global deformation
${\bf u}_\mathrm{g}$ involves a rotation and a uniform shear deformation,
smoothly transforming the rotated triangular lattice at $V=0^+$ into the
isosceles lattice locked to the substrate along the $x$-axis at large $V$.  In
our case, this isosceles triangular lattice (below called the $bb'$-lattice)
is characterized by a height $b$ (along $x$), while, in the absence of the
second substrate mode, the base $b' \approx 1.0173\, b$ along the $y$-axis can
be found from minimizing the Gibbs free energy density $g(p)$ at fixed height
$b$ and pressure $p$.

The analysis of the soliton structure in Refs.\ \onlinecite{PT_pap,PT_book}
starts from the triangular lattice at small $V$ and makes use of the
associated isotropic elastic theory. Here, we focus on the first soliton entry
into the $bb'$-lattice upon decreasing V; it then is more natural to calculate
the energy of the deformation ${\bf v}$ (defined relative to the
$bb'$-lattice) using the elastic theory of the $bb'$-lattice,
$g^\mathrm{el}_{bb'}({\bf v}) = g_p + g_\kappa + g_\mu$, with the linear term
$g_p = (\gamma_x+p)(\partial_x v_x) + (\gamma_y+p) (\partial_y v_y)$ driving
the system towards the triangular phase and $g_\kappa = \kappa_x (\partial_x
v_x)^2/2 + \kappa_y (\partial_y v_y)^2/2 + \kappa_{xy} (\partial_x
v_x)(\partial_y v_y)$ and $g_\mu = \mu_x (\partial_y v_x)^2/2 + \mu_y
(\partial_x v_y)^2/2 + \mu_{xy} (\partial_y v_x)(\partial_x v_y)$ the usual
(quadratic) elastic terms \cite{Gfed}; the coefficients are again calculated
using Ewald techniques. In this formulation, the substrate energy assumes the
simple form $e^\mathrm{sub} = (Vn'/2)[2-\cos(q v_x)]$ with $n' = 1/bb'$ the
particle density in the $bb'$-lattice.  Aligning the rotated coordinate system
$(z,z_\perp)$ with the misfit vector ${\bf p}_1$, the soliton displacement
${\bf v}(z)$ derives from a 1D sine-Gordon equation.

Using the isosceles elasticity, we find the Pokrovsky-Talapov (PT) soliton
first entering the $bb'$-lattice at $V_c^{\rm \scriptscriptstyle PT} \approx
0.0417 \, e_{\scriptscriptstyle D}$; the displacement field evolves along
$\theta \approx 45.05^\circ$ ($\theta \approx 42.13^\circ$ in the original
analysis in Refs.\ \onlinecite{PT_pap,PT_book} based on an isotropic 
elasticity, although see \cite{mistake}) and shifts the particle lattice by
${\bf d} \approx (-b, 0.70\, b)$. With decreasing substrate amplitude $V$, the
soliton density $n_{\rm\scriptscriptstyle sol}$ rapidly increases, $n_{\rm
\scriptscriptstyle sol} b \propto 1/|\ln(1-V/V_c^{\rm \scriptscriptstyle
PT})|$; the configuration with strongly overlapping solitons at small $V$ then
is equivalent to the rotated and distorted triangular phase obtained from
perturbation theory.

The soliton array obtained within the resonance approximation transforms the
$bb'$-lattice to the triangular one, while our goal here is to study the
transformation of the particle system from square to triangular.  The
solitonic instability then should appear on the background of the
period-doubled phase, which requires us to include the second harmonic of the
substrate potential into our analysis. We expect the first soliton entry in
the period-doubled phase to occur at small $V$ where we can treat the
period-doubled phase as an isosceles $bb$-lattice distorted by the relative
shift $\bar\delta = b/2-\delta$ of the two sublattices. Inside the soliton,
the amplitude of this short-scale distortion $\bar\delta = (b/\pi)
\arcsin(V\cos(qv_y) /8\Delta)$ is slaved to the center of mass coordinate
${\bf v(R)}$ replacing the scalar variable $\sigma$ introduced
above.  We then have to minimize the energy
\begin{eqnarray} \label{eq:g_ca2m}
   \delta g &=&\frac{1}{N}\!\int\!\! d^2R\, \Bigl\{g^\mathrm{el}_{bb}({\bf v})
    +\frac{Vn}{2}[1-\cos(q v_x)]
    \\ \nonumber
    &&\qquad\qquad\qquad + \frac{V^2n}{64 \Delta}[1-\cos(2qv_y)]\Bigr\},
\end{eqnarray}
where $g^\mathrm{el}_{bb}$ is the elastic Gibbs free energy \cite{Gfed}
density of the $bb$-lattice.  While the resonance approximation admits only
one low-energy soliton, the full problem with both substrate modes present
allows for several line-defects shifting the lattice by ${\bf d}_{j,k} =
(-jb,kb/2)$ with $j,k$ integers. Promising candidates reminding about the PT
soliton are the $(j,k) = (1,k)$ defects, but a simple Ansatz with the shift
${\bf d}_{01} = (0,b/2)$ should be tried as well, since the particles merely
have to overcome the weak effective potential $\propto V^2/64\Delta \ll V/2$
along the $y$-direction, see Eq.\ (\ref{eq:g_ca2m}).  All these line defects
fall into two classes, the domain walls with $j+k$ assuming odd values and
taking the period-doubled phase from one twin to the other, $\delta \to
-\delta$, and the genuine solitons with $j+k$ even and the same twin on
both sides, $\delta \to \delta$.

\begin{figure}[h]
\begin{center}
\includegraphics[width=7cm]{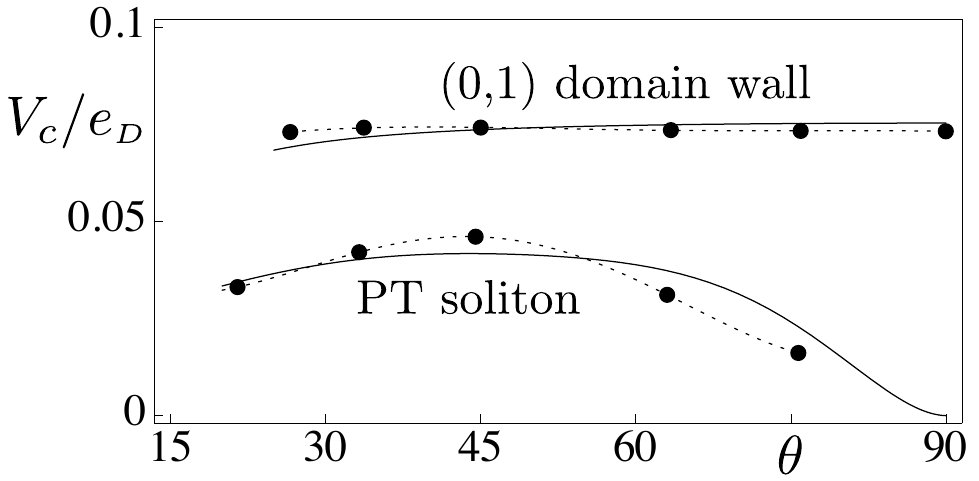}
\end{center}
\caption{\label{fig:num_res} 
Numerical results for the critical substrate potential $V_c$ for first soliton
entry versus angle $\theta$. Shown is the data for the $(0,1)$ domain wall and
for the PT soliton evaluated at selected angles defined by small Miller
indices $(m,n)$; dotted lines are guides to the eye.  The flat form
$V_c^{\scriptscriptstyle (0,1)}(\theta)$ renders the angle $\theta$ for the
first $(0,1)$ domain wall entry poorly defined.  Thin lines are the analytic
results following from a continuum-elastic description for an isosceles
lattice.}
\end{figure}

The determination of the critical substrate potential for the $(0,1)$ domain
walls is straightforward,
\begin{eqnarray} \label{eq:V_c^v}
   V_c^{\scriptscriptstyle (0,1)} = -\frac{2\pi(\gamma_y+p)}{n}
        \sqrt{\frac{n \, \Delta} {\kappa_y+\mu_y\cot^2\theta}},
\end{eqnarray}
and provides the maximal value $V_c^{\scriptscriptstyle (0,1)} \approx 0.0753
\, e_{\scriptscriptstyle D} > V_c^{\rm \scriptscriptstyle PT}$ at $\theta =
90^\circ$, see Fig.\ \ref{fig:num_res}.  The analysis for the
$(1,k)$ defects is more involved and the results depend strongly on the type
of elasticity theory chosen for the calculation,
telling us that corrections due to anharmonicities are large.

For this reason, a reliable conclusion on the relevant scenario requires an
numerically precise computation of the defects' Gibbs free energies.  Starting
from a variational Ansatz, we relax the particle configuration numerically for
periodic arrays with large separations between the defects.  Summing up terms
along the direction perpendicular to $z$ reduces the problem to a 1D one, but
restricts the possible angles $\theta$ to those appertaining to small Miller
indices.  The results for the $(0,1)$ domain wall (extrapolated to the
thermodynamic limit) are shown in Fig.\ \ref{fig:num_res}; they agree well
with the analytic ones, although the largest $V_c^{\scriptscriptstyle (0,1)}
\approx 0.07415 \, e_{\scriptscriptstyle D}$ is assumed at a different angle
$\theta = 45^\circ$.  While the flat dependence on angle renders the optimal
orientation of the domain wall poorly defined, the data shows that the optimal
defect does not align with a symmetry axis of the isosceles lattice. This
result is quite unexpected, as such a symmetry alignement is predicted by the
analytic calculation neglecting anharmonicities and has often been considered
as natural in the literature \cite{Chaikin_95}.  Our numerical results
\cite{Barbara_13} for the $(1,k)$ defects show that these would appear at much
smaller values of $V$; in particular, the (second) best result
$V_c^{\scriptscriptstyle (1,3)}(\theta = 45^\circ) \approx 0.0544 \,
e_{\scriptscriptstyle D}$ is found for the $(1,3)$ soliton, while the $k=2$
domain wall and $k=1$ soliton are even worse with $V_c^{\scriptscriptstyle
(1,2)}(45^\circ) \approx 0.0501 \, e_{\scriptscriptstyle D}$ and
$V_c^{\scriptscriptstyle (1,1)}(63^\circ) \approx 0.0382 \,
e_{\scriptscriptstyle D}$.

The proliferation of $(0,1)$ domain walls washes out the $y$-harmonic and
dilutes the particles along the $y$-axis, thereby establishing the
$bb'$-lattice; the transformation to the (rotated) triangular lattice at $V =
0^+$ then involves an additional PT solitonic transition at lower $V$ which
smoothly eliminates the $x$-harmonic.  The analytic result for $V_c^{\rm
\scriptscriptstyle PT}$ again can be improved with a numerical calculation and
we find a maximal critical potential $V_c^{\rm \scriptscriptstyle PT}(\theta
\approx 44.5^\circ) \approx 0.046 \, e_{\scriptscriptstyle D}$, see Fig.\
\ref{fig:num_res}; at this value of the substrate potential, the domain-wall
phase has approached the $bb'$-lattice to within $\approx 10$ \%, as measured
by the ratio of amplitudes $A_\mathrm{p}$ of the periodic deformation ${\bf
v}_\mathrm{p}$ generated by the $(0,1)$ domain wall array, $A_\mathrm{p}
(V_c^{\rm\scriptscriptstyle PT})/A_\mathrm{p} (V_c^{\scriptscriptstyle (0,1)})
= 0.019/0.25 \approx 0.08$.

Depending on the specific situation at hand, alternative scenarios can be
realized. All of these have to respect that a phase transition establishing an
array of identical solitons with shift vector ${\bf d}$ [e.g., $(1,k)$
solitons or domain walls] necessarily has to be followed by a further
transition at lower $V$; since the global distortion field in the soliton
array is slaved to ${\bf d}$, the rotated triangular phase at $V = 0^+$ cannot
be reached without the appearance of other defects.  The completion of the
transformation may then involve the formation of a network of crossing
solitons. Furthermore, if the most favorable solitons have close critical
potentials and intersect with a negative energy, the two smooth transitions
can merge into a single first-order one.

To conclude, we discuss the prospects for an experimental realization and
detection of these competing structures in a cold molecule system.  In order
to serve as a classical simulator, quantum fluctuations have to remain small.
While in usual cold atom systems the latter are limited by the optical
lattice, here it is the long-range interaction between the molecules that
bound the zero-point motion. In estimating the importance of quantum
fluctuations, we have to compare the interaction energy $e_{\scriptscriptstyle
D}$ with the recoil energy $e_\mathrm{r} = \hbar^2/mb^2$. Evaluating the
quantum parameter $r_Q = e_{\scriptscriptstyle D}/e_\mathrm{r} \approx 15\, Z
D [\mathrm{D}^2]/b[\mathrm{nm}]$ (with $Z$ denoting the molecular
mass and $\mathrm{D}$ the Debye unit) for favorable but reasonable parameter
settings ($Z \sim 100$, $b \sim 500$ nm, $\sqrt{D} \sim 5~\mathrm{D}$), we
obtain $r_Q \sim 10^2$. This is substantially larger than the critical value
$r_Q = r_\mathrm{sf} \approx 18$ \cite{Buechler_07} marking the transition to
the superfluid state where quantum fluctuations dominate \cite{limit}.  Hence
molecular systems can serve as classical simulators, although some
renormalization effects due to quantum fluctuations may occur. Furthermore,
sufficiently large amplitudes $V$ must be reached for the optical lattice; in
a recent experiment \cite{Chiota_12}, dipolar molecules have been localized in
deep wells $V \sim 10^2 \, e_\mathrm{r}$, which should be sufficient to reach
the critical value $V_{\scriptscriptstyle \square}$.  Finally, a promising way
to identify the various structural phases is via their different dynamical
response under an applied force field ${\bf f}$, with the square and
period-doubled phases characterized by symmetric and asymmetric (reduced along
$y$) pinning, respectively. For practical purposes, the exponentially weak
pinning of solitons can be neglected; the force field ${\bf f}$ induces a
drive ${\bf f}\cdot {\bf d}_{j,k}$ along $z$ and the resulting soliton motion
generates a mass flow along ${\bf d}_{j,k}$ which allows to identify the two
solitonic phases.
%
%
%

We thank Hanspeter B\"uchler, Tilman Esslinger, Sebastian Huber, Matthias
Troyer, and Thomas Uehlinger for helpful discussions and acknowledge financial
support of the Fonds National Suisse through the NCCR MaNEP; one of us (SEK)
thanks the Pauli Center for Theoretical Physics for its generous hospitality.


\begin{thebibliography}{10}
\bibitem{mistake} In Ref.\ \onlinecite{PT_pap} the Poisson ratio was
mistakenly evaluated to $\nu = 5/11$ producing an angle $\theta \approx
24^\circ$ (corrected in Ref.\ \onlinecite{PT_book}).
\bibitem{limit} Exploiting the large dipole moment $\sqrt{D} = 8.9$ D
\cite{Kaufmann_65} of SrO and using $b \sim 300$ nm one can reach $r_Q \sim
400$ \cite{Buechler_07}.
\end{thebibliography}
\end{document}